\documentclass[a4paper,11pt]{article}
\usepackage{pos}

\usepackage{siunitx}
\usepackage{placeins}
\usepackage{graphicx}
\usepackage{caption}
\usepackage{subcaption}
\usepackage{tikz}

\title{Hybrid cosmic ray measurements using the IceAct telescopes in coincidence with the IceCube and IceTop detectors}
\ShortTitle{Hybrid cosmic ray measurements using the IceAct telescopes}
\author{The IceCube Collaboration \\{\normalsize \normalfont(a complete list of authors can be found at the end of the proceedings)}}

\emailAdd{larissa.paul@icecube.wisc.edu}
\emailAdd{matthias.plum@icecube.wisc.edu}
\emailAdd{merlin.schaufel@icecube.wisc.edu}

\abstract{IceAct is a proposed surface array of compact (50 cm diameter) and cost-effective Imaging Air Cherenkov Telescopes installed at the site of the IceCube Neutrino Observatory at the geographic South Pole.  Since January 2019, two IceAct telescope demonstrators, featuring 61 silicon photomultiplier (SiPM) pixels have been taking data in the center of the IceTop surface array during the austral winter. We present the first analysis of hybrid cosmic ray events detected by the IceAct imaging air-Cherenkov telescopes in coincidence with the IceCube Neutrino Observatory, including the IceTop surface array and the IceCube in-ice array. By featuring an energy threshold of about 10 TeV and a wide field-of-view, the IceAct telescopes show promising capabilities of improving current cosmic ray composition studies: measuring the Cherenkov light emissions in the atmosphere adds new information about the shower development not accessible with the current detectors, enabling significantly better primary particle type discrimination on a statistical basis. The hybrid measurement also allows for detailed feasibility studies of detector cross-calibration and of cosmic ray veto capabilities for neutrino analyses. We present the performance of the telescopes, the results from the analysis of two years of data, and an outlook of a hybrid simulation for a future telescope array.

\vspace{4mm}
{\bfseries Corresponding authors:}
Thomas Bretz$^{2}$, Giang Do$^{2}$, John W. Hewitt$^{3}$, Frank Maslowski$^{2}$, Larissa Paul$^{1}$, Matthias Plum$^{1}$, Florian Rehbein$^{2}$, Johannes Sch\"afer$^{4}$, Merlin Schaufel$^{2*}$, Adrian Zink$^{4}$\\
{$^{1}$ \itshape Marquette University, Milwaukee,WI USA}\\
{$^{2}$ \itshape RWTH Aachen University, Aachen, Germany}\\
{$^{3}$ \itshape University of Northern Florida, Jacksonville,FL USA}\\
{$^{4}$ \itshape Erlangen Centre for Astroparticle Physics, Erlangen, Germany}\\[4mm]
$^*$ Presenter

\FullConference{37$^{\rm{th}}$ International Cosmic Ray Conference (ICRC 2021)\\
		July 12th -- 23rd, 2021\\
		Online -- Berlin, Germany}

}

\begin{document}
\maketitle
\section{Introduction}

IceCube \cite{Aartsen:2016nxy} is a cubic-kilometer neutrino detector installed in the ice at the geographic South Pole between depths of 1450 m and 2450 m, completed in 2010. Reconstruction of the direction, energy, and flavor of the neutrinos relies on the optical detection of Cherenkov radiation emitted by charged particles produced in the interactions of neutrinos in the surrounding ice or the nearby bedrock.
Further, IceCube and its surface component IceTop \cite{icetop2013} provide a unique environment for the study of cosmic rays. The enhancement of the surface component with additional detectors like imaging air-Cherenkov telescopes (IACTs \cite{iceact_demonstrator,Schaufel:2019aef}, scintillators, and radio antennas \cite{Kauer:2019K+,marie_roxanne} will improve these capabilities even further. 

As a proposed array of compact IACTs, the IceAct telescopes feature a much lower energy threshold compared to the other surface components. The combination of these detectors forms a hybrid detector providing the information from the electromagnetic shower development in the atmosphere (IceAct, Radio Array), the electromagnetic and muonic distribution on the surface (IceTop, Scintillator Array), and the high energetic muon bundles (IceCube). The hybrid detection results in a better particle type identification, better geometrical reconstruction, ability of cross-checking the energy scale of different detector types, calibrating possibilities of the directional uncertainty of the in-ice muon reconstruction, and potentially enhanced veto capabilities of low energy cosmic ray air  events (above 50\,TeV) for neutrino analyzes. As an example, the energy reconstruction for a single IACT is very challenging as the measured brightness, which is the most energy-sensitive variable, depends not only on the primary particle energy but also on the distance between the shower impact point on the surface and the telescope. Combining the geometric distance information from IceCube with the image size (summed brightness of an image) measured by IceAct solves this ambiguity between low energetic near showers and high energetic distant showers. 


\section{Setup and performance}

To achieve its science goals, the IceAct telescope is designed to withstand the harsh conditions at the South Pole and to be able to operate continuously during the Antarctic winter \cite{iceact_demonstrator}. The telescope is build from a lightweight fiberglass barrel hosting a 55\,cm wide UV transparent Fresnel lens and the camera with integrated data acquisition (DAQ) system \cite{Bretz_2018}. The camera consists of 61 pixels based on 6\,mm silicon photomultiplier (SiPM) SensL-FJ-60035 each equipped with a solid PMMA light-guide. The telescope has a focal length of 50.2\,cm, a single pixel has a field-of-view (FOV) of approximately 1.5\,$^\circ$ compiling to a field-of-view of 12\,$^\circ$ for a telescope. To keep the lens free of snow and ice, it is protected by a thin glass-plate which can be heated using a coil of a self-regulating heating wire between the lens and the protective plate. The heating wire is held by an aluminum structure to ensure an even heat spread and to prevent any force on the lens. For data acquisition, a \textit{TeV Array Readout Electronics with GSa/s sampling and Event Trigger} (TARGET) module is used \cite{doi:10.1063/1.4969033}. It provides 64 channels for digitization and triggering, as well as an integrated SiPM bias supply for 16 channels. Therefore, the SiPMs are organized in groups of 4 to share a bias voltage and form a trigger group.
To compensate the temperature dependence of the SiPM the bias supply measures the temperature of the camera and corrects the bias voltage accordingly.

\begin{figure}
	\centering
	\includegraphics[width=0.99\textwidth]{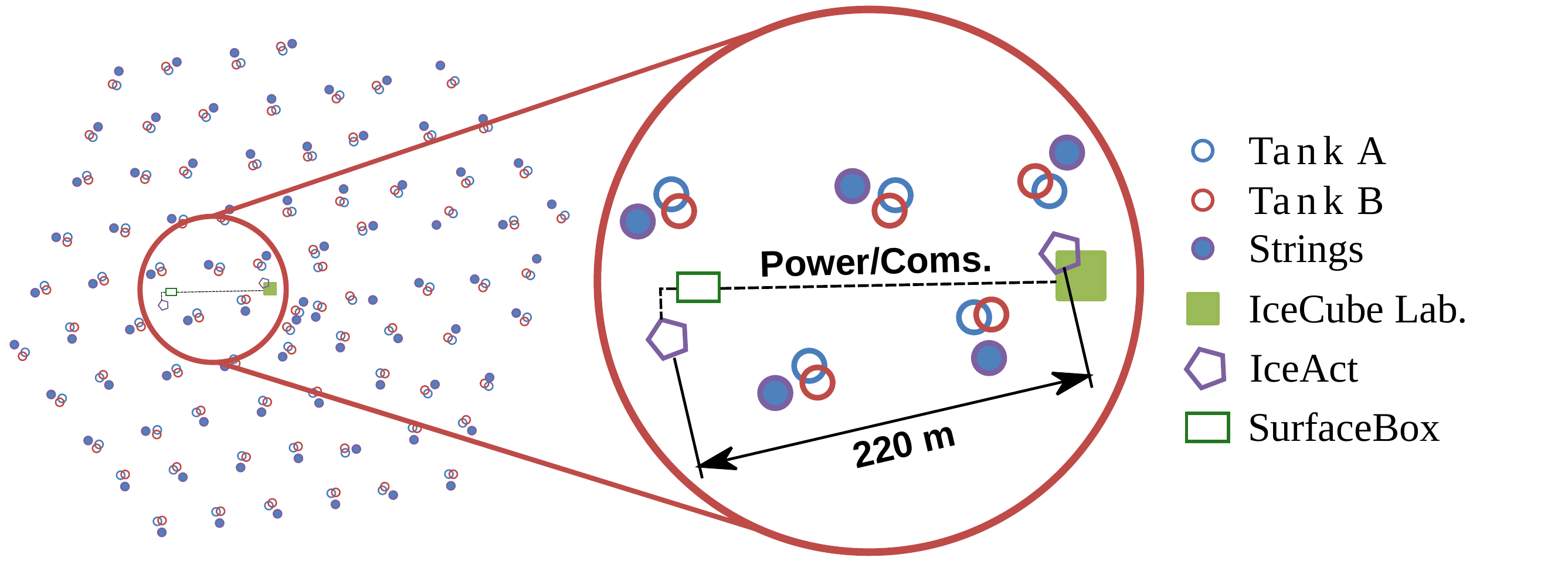}
	\caption{\label{Plt:Sketch}Overview of the IceAct setup since 2019: One IceAct telescope is mounted on the roof of the IceCube Laboratory in the center of the array (referred to as the Roof telescope). A second telescope is installed on the ice (referred to as the Field telescope). The distance is approximately 220\,m.}
\end{figure}

Since 2019 two IceAct telescopes operate at the South Pole. One is mounted on the roof of the IceCube laboratory (referred to as the Roof telescope) and one is installed on an aluminum stand in a distance of 220\,m (referred to as the Field telescope). Figure \ref{Plt:Sketch} shows the position of the two telescopes in the IceTop/IceCube array. 
Studies with telescope prototypes are showing that Aurora and Moon light have an impact on the event trigger rate, but not the overall image/event quality. The current engineering setup showed that these telescope can achieve a duty cycle of more than 20$\%$ average over the full year in the last years.

\section{Coincident events}

To analyze hybrid events, coincident events from the IceCube and the IceAct data streams have to be synchronized. For an event-by-event matching, the synchronization needs to be precise enough to ensure a low rate of random coincidences. The matching is currently done differently for the Roof and the Field telescope. The Roof telescope has a trigger signal connected directly to the IceCube DAQ system \cite{IceCube:2008qbc}, capable to flag coincident events in the IceCube data stream or even to trigger the event builder to save all coincident IceCube and IceTop hits. The time of the IceAct Roof trigger is then automatically saved in the IceCube GPS time. 

As the Field telescope does not have any direct trigger connection, the synchronization uses the internal trigger timestamps. The IceAct DAQ stores the timestamp with a 1\,ns resolution. The limiting factor of the internal timestamp is the drift of the 125\,MHz oscillator, which can be corrected to provide a timestamp with a RMS better than 30\,$\mu$s compared to the IceCube GPS time. This time uncertainty already includes IceCube trigger time jitter and other physical timing effects, like muon path length differences for different zenith angles etc.\ and is sufficient for an event to event matching. 

For in-event timing, the trigger time needs to be known with a nanosecond precision. For the Roof, this is achieved by storing the trigger time directly in the IceCube data stream. The Field telescope triggers are additionally timestamped using a White Rabbit \cite{wrproject} based trigger module which is synchronized to the main IceCube GPS clock with a sub-nanosecond precision. The module saves a precise timestamp for every trigger of the connected telescope and is foreseen as trigger module for a station consisting of seven IceAct telescopes. 


To demonstrate the successful matching of coincident events and the capabilities of the 61 pixel telescopes a first analysis of a single day of data is presented in the following. 
The data was taken on the 27th of July 2019 and serves as test sample to develop and test analysis methods for the 2019 dataset. The selection of the test samples is based on good run conditions: Constant trigger rate in both telescopes, low Aurora activity and clear sky.  

\begin{figure}[htb]
	\centering
	\begin{subfigure}[b]{0.5\textwidth}
		\centering
		\includegraphics[width=\textwidth]{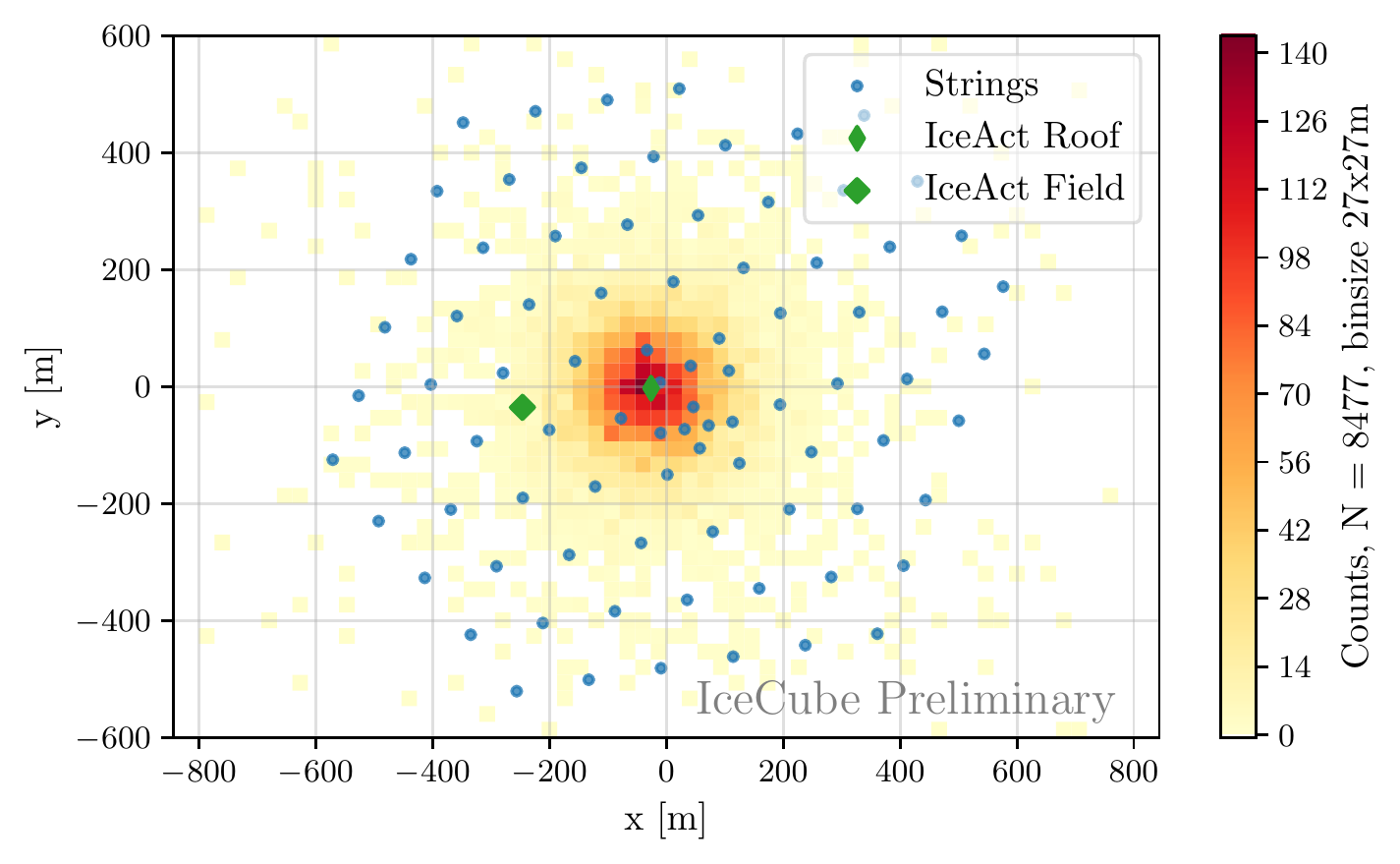}
	\end{subfigure}
	\hfill
	\begin{subfigure}[b]{0.48\textwidth}
		\centering		
		\includegraphics[width=\textwidth]{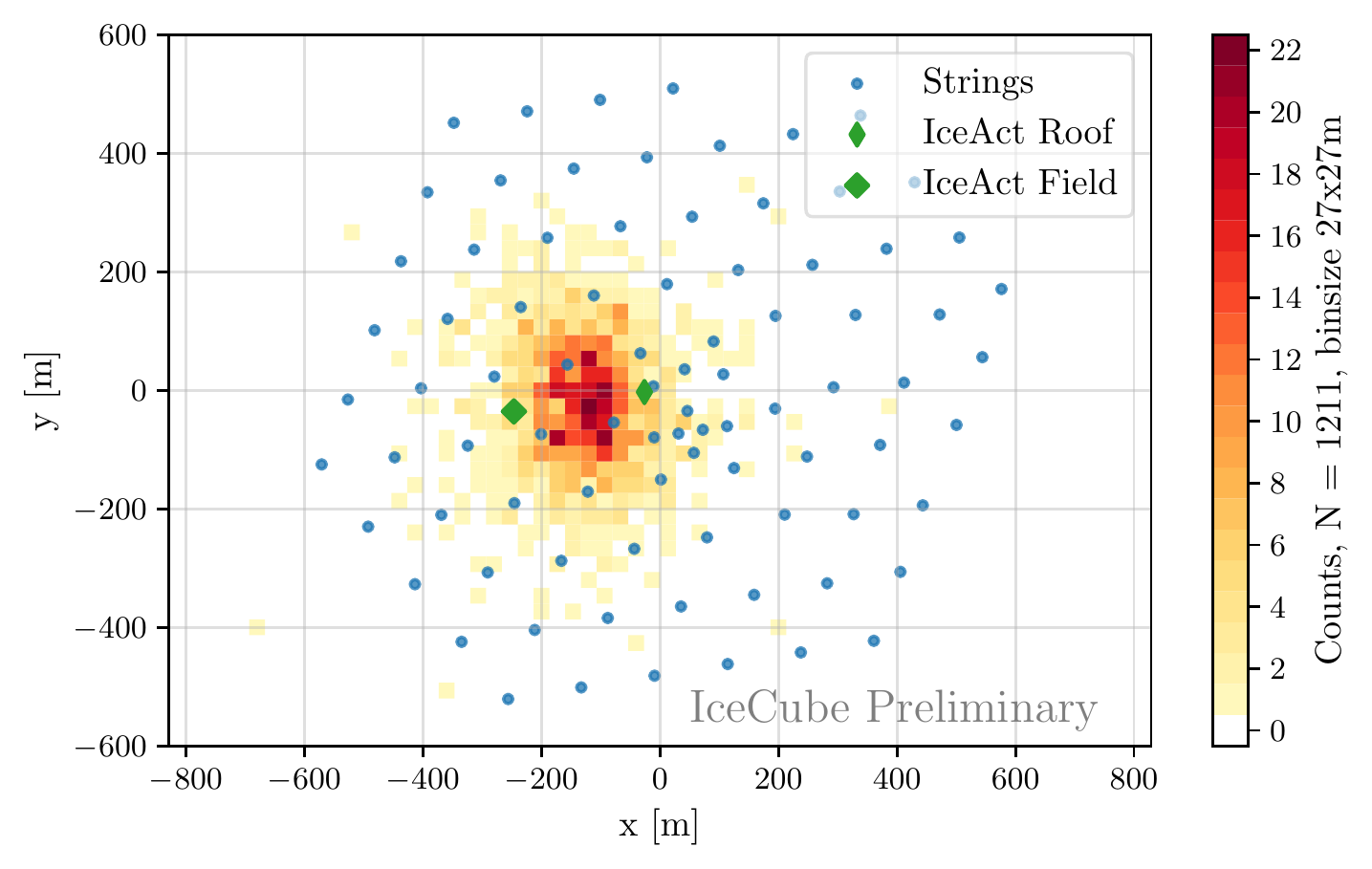}
	\end{subfigure}
	\caption{\label{Plt:Dist} Comparison of surface positions of coincident and reconstructed muons \cite{Ahrens:2003fg,AARTSEN2014143} with mono (left) or stereo (right) IceAct trigger for one day of data taking in 2019.}
\end{figure}

For coincident events the reconstruction of muons reaching the in-ice detector is extrapolated back to the surface. This surface positions of coincident muons from events triggered by the Roof telescope are shown in Figure \ref{Plt:Dist} (left). One can clearly see that the events are clustering around the position of the Roof telescope, as expected due to the lower light yield of a more distant shower of the same energy. In Figure \ref{Plt:Dist} (right) the distribution of extrapolated positions of in-ice muons at the surface are shown for events where both telescopes and IceCube have triggered (referred to as stereo events). 
The stereo events are distributed symmetrically between both telescopes.


As Cherenkov emission from extensive air showers is strongly beamed towards the shower axis the arrival direction of this light is weakly correlated with the direction of the shower (limited by the opening angle of the Cherenkov cone). The telescopes imaging lens translates this direction into a position on the camera plane. Therefore, the center of gravity of the telescope image provides direct information on the shower direction. Even with such a simple direction estimator, the 61-pixel images provides a rough approximation of the shower direction.

To compare these directions to the reconstructed muon directions \cite{Ahrens:2003fg,AARTSEN2014143}, coincident events are cleaned, to exclude non-coincident signals to minimize the noise content in the camera image. After image-cleaning the COG is calculated and a simple direction estimator is derived as:

\[\Phi = \tan^{-1}\left(\frac{Y_{\mathrm{COG}}}{X_{\mathrm{COG}}}\right) \hspace{1cm} \Theta = \tan^{-1}\left(\frac{\sqrt{X_{\mathrm{COG}}^{2}+Y_{\mathrm{COG}}^{2}}}{f_{\mathrm{len}}}\right)\]

With $X_{\mathrm{COG}}$ and $Y_{\mathrm{COG}}$ being the $X$ and $Y$ position of the image center of gravity and $f_{\mathrm{len}}$ the focus length of the fresnel lens.

\begin{figure}[htb]
	\centering
	\begin{subfigure}[b]{0.47\textwidth}
		\centering
		\includegraphics[width=\textwidth]{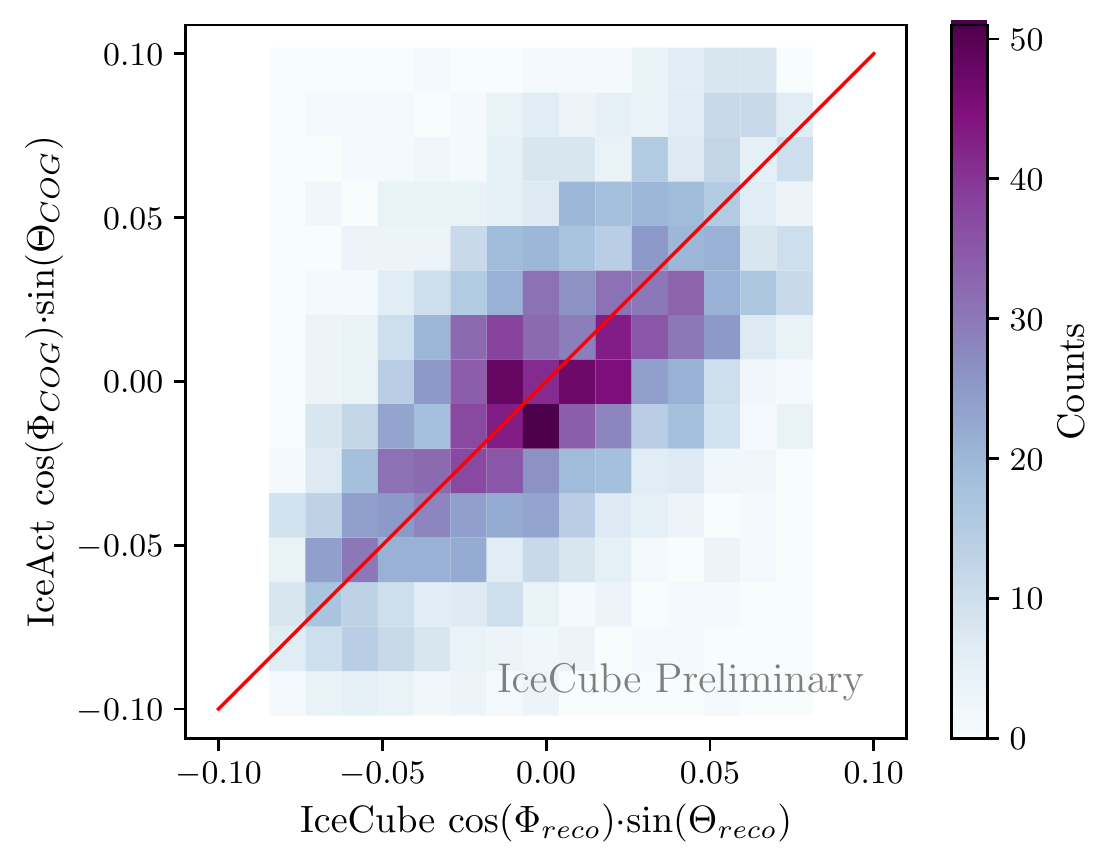}
	\end{subfigure}
	\hfill
	\begin{subfigure}[b]{0.47\textwidth}
		\centering		
		\includegraphics[width=\textwidth]{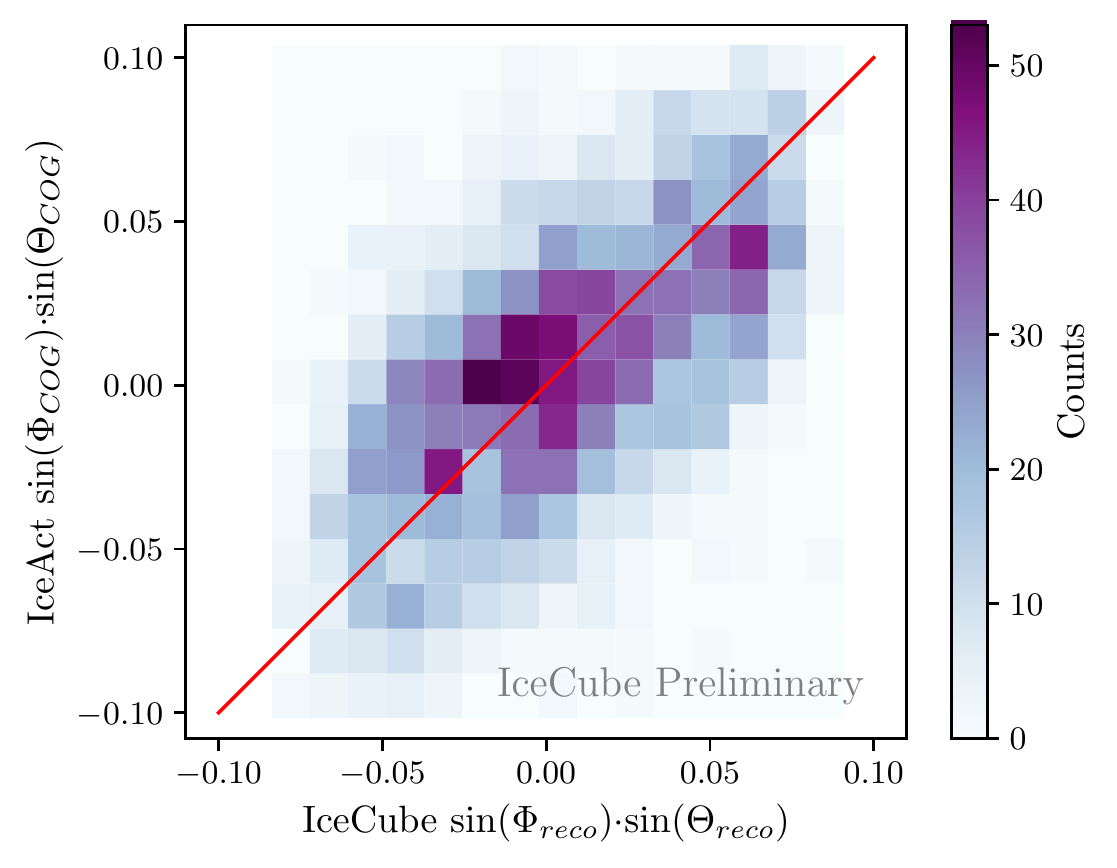}
	\end{subfigure}
	\caption{\label{Plt:UVCorr} Direction correlation of coincident events reconstructed by IceCube (level 2 log-likelihood reconstruction) and IceAct (direction of the image center-of-gravity after image cleaning). Shown are the X-projection (left) and the Y-projection (right) of the directional vectors on ground level.}
\end{figure}

To test the correlation we used Roof telescope data from the 2019 test sample. As the Roof telescope is rotated and tilted against the IceCube reconstruction coordination system, all IceCube events are rotated by $(204.78^{\circ}$,$-0.97^{\circ}$) in azimuth and zenith to match the direction of the telescope axis. The directions are represented as vectors and any rotation applied to the events is done in 3D. This rotation is determined by minimizing the mean angular distance between the IceCube and IceAct events and is limited to the assumption of no large systematic shifts in the IceCube reconstruction. The correlation of the IceCube reconstruction and the IceAct directional proxy can be seen in Figure \ref{Plt:UVCorr}. To display the directional correlation, the $X$ and $Y$ projection of the direction vector are shown. Even with the simple directional proxy one can see a clear correlation with a spread of roughly $2^{\circ}$. This is well compatible with the expectation for the Level 2 Muon reconstruction and the simple telescope direction proxy. 

\section{Simulation and reconstruction}
To evaluate the combined hybrid detector in-ice IceCube, IceTop, and IceAct a joined simulation dataset is used and a simple analysis of the IceAct events is applied and compared to a test sample from early August 2020. 
Air shower in the atmosphere are simulated by CORSIKA \cite{HeckKnappCapdevielle1998_270043064}, with FLUKA \cite{FLUKA:2006} as the low-energy hadronic interaction model and SIBYLL-2.3c \cite{riehn2017hadronic} as the high-energy interaction model.  
The air showers are processed in the Icetray \cite{icetray} software framework to convert the particle information into a trigger/detector response. Standard IceCube event filtering and reconstruction are applied to the IceCube \cite{Aartsen_2014} and IceTop \cite{icetop2013} events. The IceAct events are also processed in the IceCube software with a full electronic/trigger simulation including a night sky background rate of 39\,MHz. 
The protons, helium, nitrogen, aluminium, and iron data sets are simulated with a differential $E^{-1}$ spectrum between 3$\,$TeV and 1$\,$PeV up to a zenith angle of 20$^\circ$ for the full azimuth range. The core position of the shower is scrambled inside a circle with increasing radius for increasing energy starting at 250$\,$m for energies above 3$\,$TeV. The generator assumes a layered parameterization of the South Pole atmosphere based on average April atmospheres \cite{sam_thesis}.

The composition sensitivity in this energy range is studied, using a simple principle component analysis of the center-of-gravity and the spread of the light distribution \cite{1985hillas} for the simulated IceAct events. These Hillas parameters are the total light distribution (\textit{image size}) and the standard deviation along the major and minor image axis (\textit{length} and \textit{width}).
The quality of the IceAct events is ensured by a simple cut on the reconstructed peak height as image cleaning and requiring that each event has at least 3 surviving reconstructed peak heights to calculate the Hillas parameters.

\begin{figure}
	\centering
	\subfloat{\includegraphics[width=0.5\textwidth]{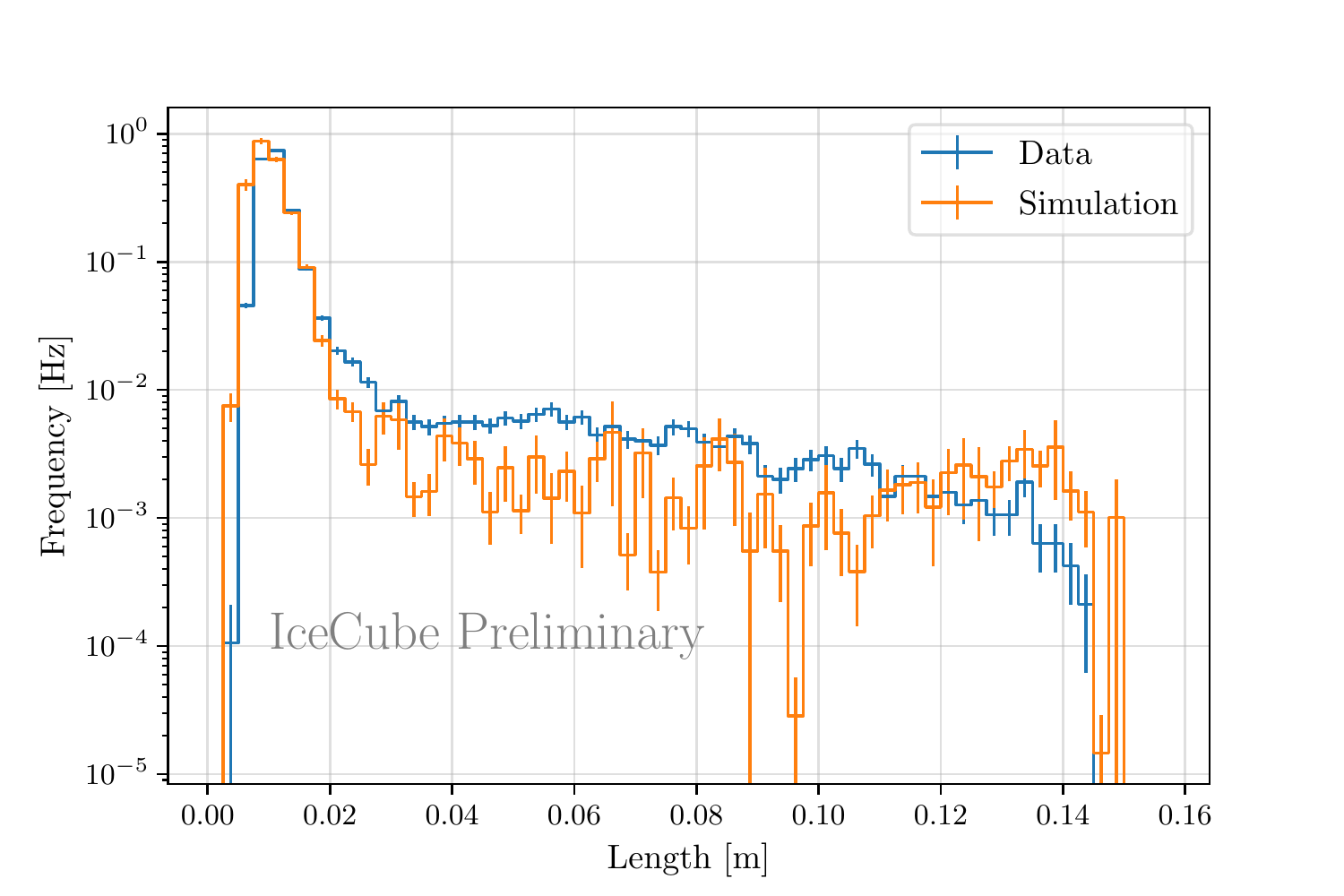}}
	\subfloat{\includegraphics[width=0.5\textwidth]{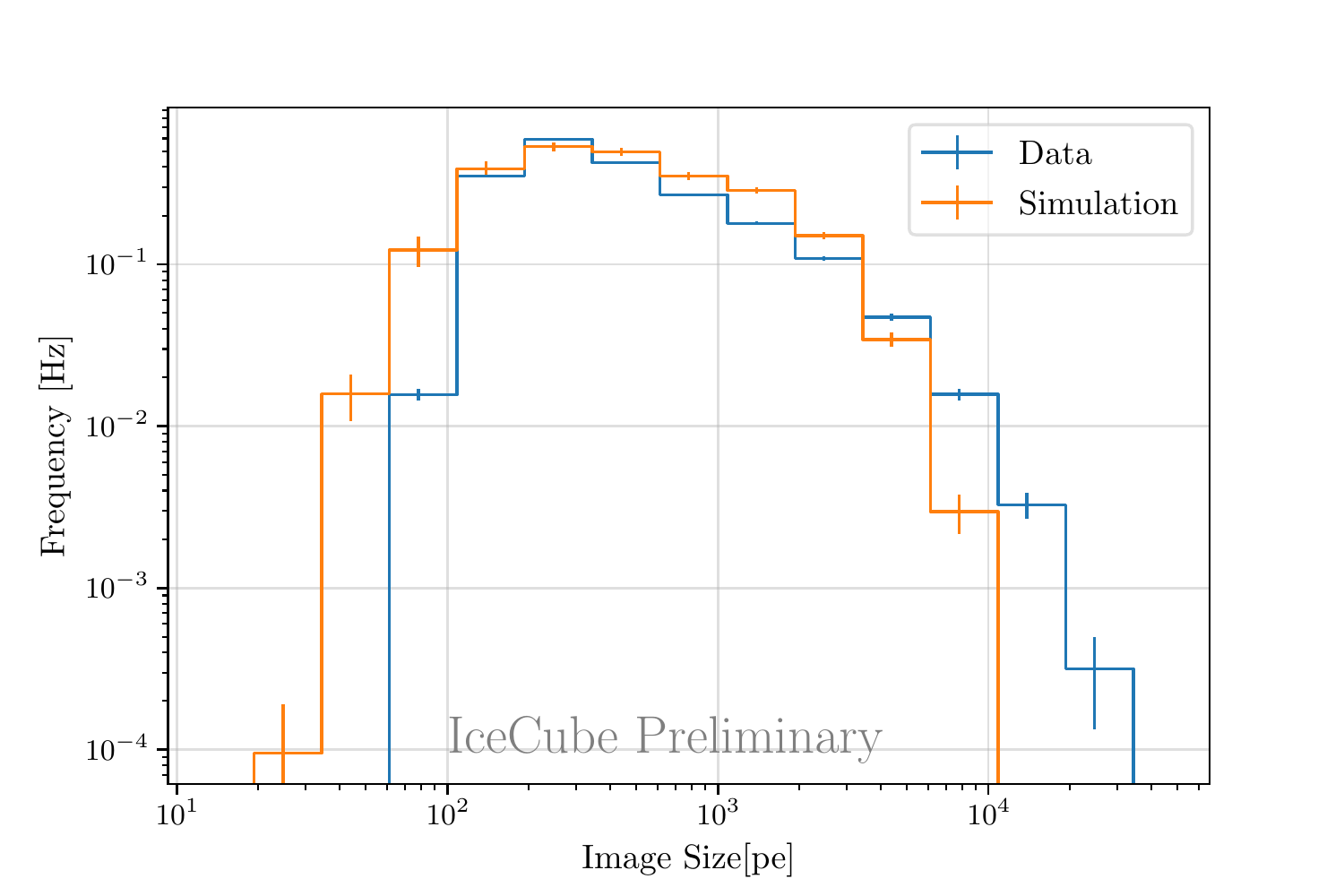}}
	\caption{\label{Plt:data_MC_hillas} Data/Monte-Carlo agreement for Hillas image parameters \cite{1985hillas}. (left) the standard deviation along the major image axis (\textit{length}) in pixel coordinates in meter (right) calibrated total light distribution (\textit{image size}) in photo electrons (pe)}
\end{figure}

A first data/MC agreement study of the Hillas parameter \textit{length} and \textit{image size} is shown in Figure \ref{Plt:data_MC_hillas} reweighting the MC to an H4a \cite{Gaisser_H4a} cosmic ray flux. 
The data/MC distributions show the same underlying shape in these parameters. The peak in the \textit{length} is clearly visible in both distributions and the tail has a similar shape, which is slightly shifted. The \textit{image size} shows some discrepancy at the upper end of the distributions is probably due to the limited energy range of the MC to only 1 PeV whereas the data contains also higher energies. The calibration of the telescopes and the simulation chain are still in development and are currently being improved.

\begin{figure}
	\centering
	\subfloat{\includegraphics[width=0.5\textwidth]{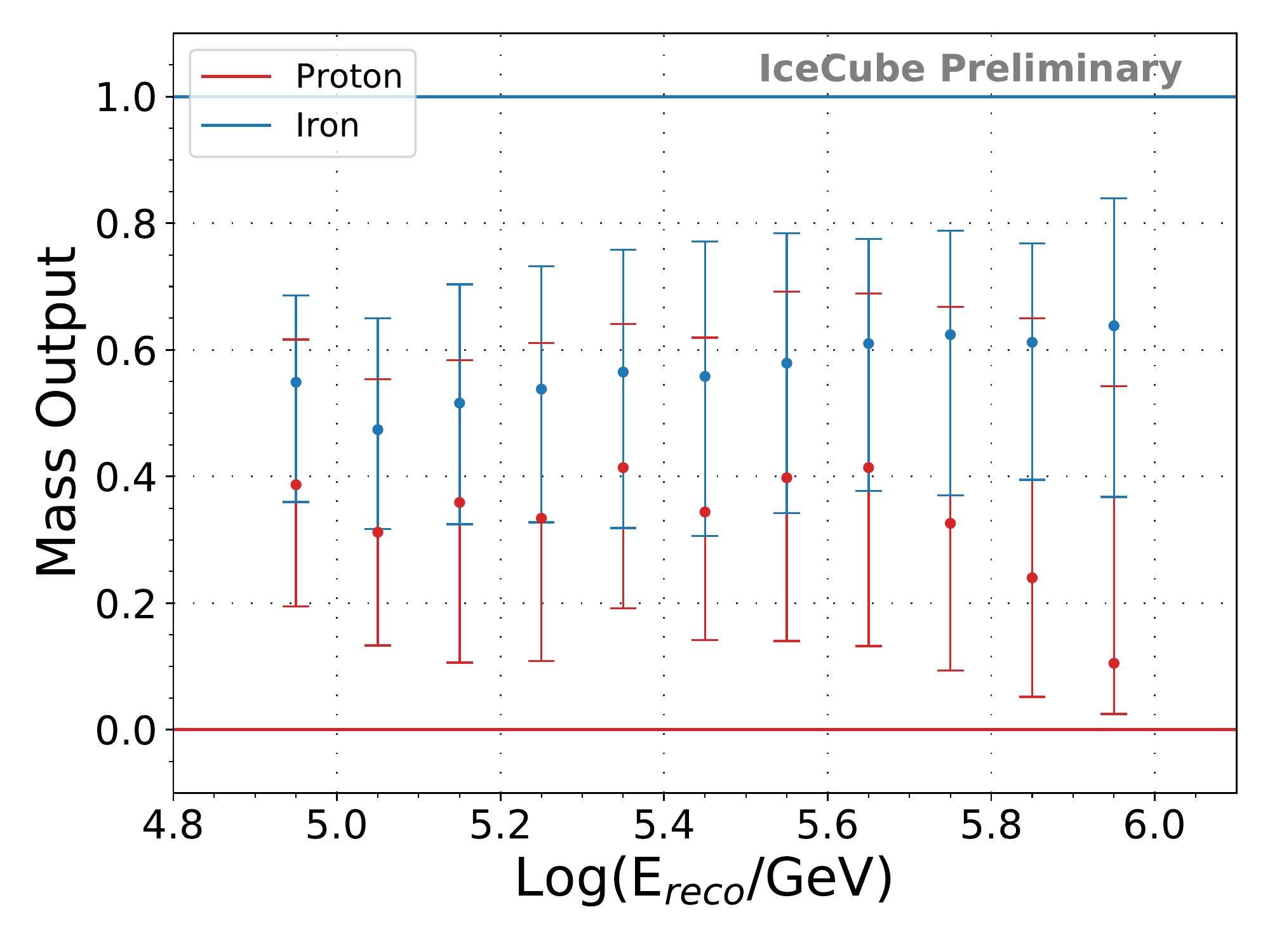}}
	\subfloat{\includegraphics[width=0.5\textwidth]{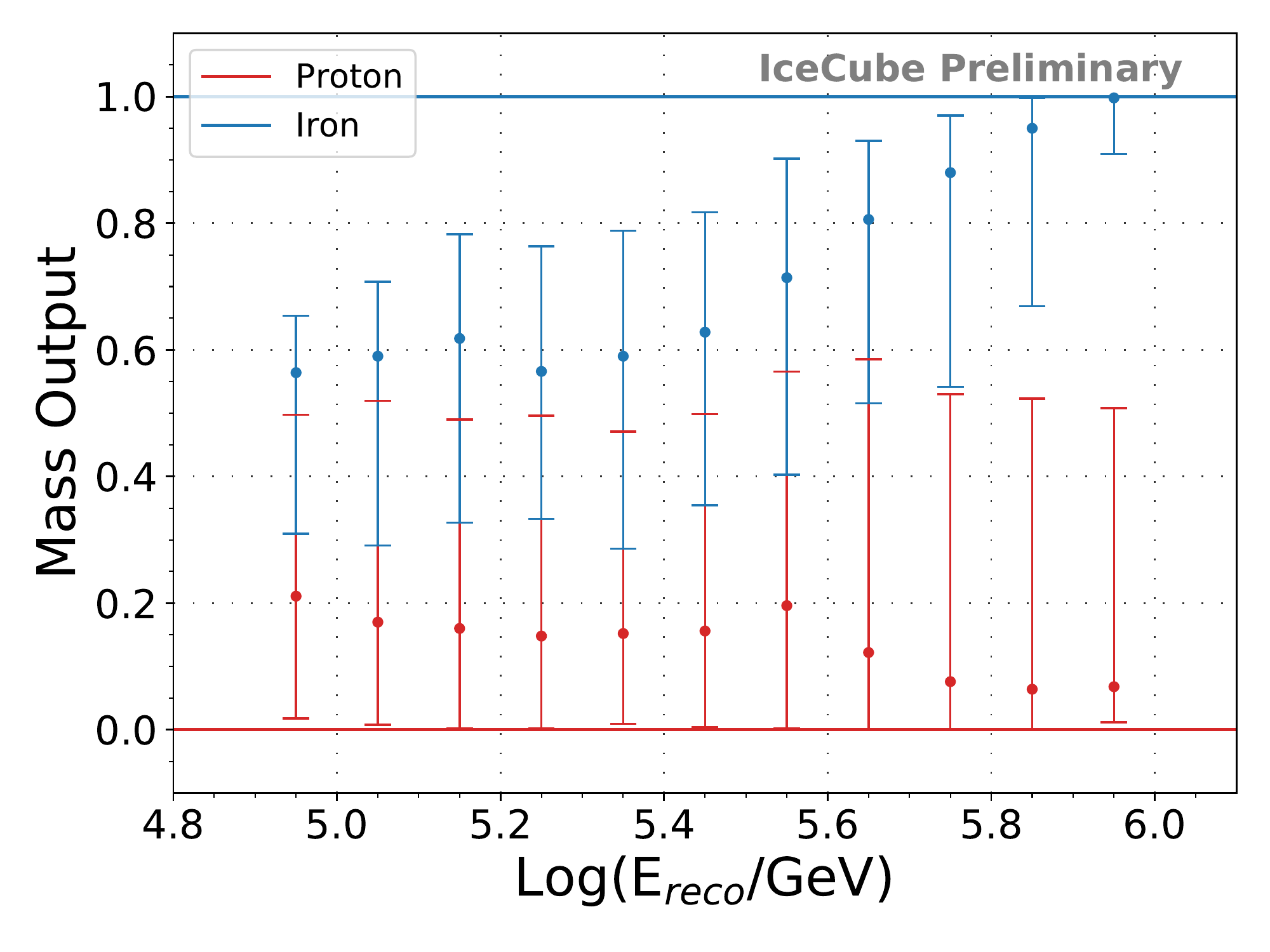}}
	\caption{\label{Plt:RFReco}
	(left) Median and 68\% quantile of RF mass output as a function of the reconstructed energy for a single IceAct telescope events. (right) Median and 68\% quantile of RF mass output as a function of the reconstructed energy for hybrid IceAct/IceCube events.}
\end{figure}

Using only the IceAct telescopes in a MC study, the Hillas image parameters, the distance to the shower core and zenith are analysed with a Random Forest regression analysis (similar to \cite{paras_plum_saffer}) to reconstruct the primary energy and mass. The total number of trees in the ensemble is set to 500 with a maximum depth of 150. The tree split criterion is the mean square error. A bootstrap method is applied and an out-of-bag sample to estimate the generalization score. All other values are kept to the default values in \cite{scikit-learn}.
The training uses 66\%-of the MC and the rest is used to test the energy and mass reconstruction.
The results of the mass reconstruction is shown in Figure \ref{Plt:RFReco} (left plot), where the median and the 68\%-quantile for different primaries are shown as a function of the reconstructed energy.
This figure shows a small reconstruction bias for the mass output, however as it is shown in \cite{composition_paper}, a template-style analysis can produce mass composition results with good accuracy even in case of these kinds of biases. 

The analysis can be improved by using hybrid event information, e.g.\ the high energy muon response, which is measure by the in-ice IceCube array. This additional information is providing an improvement in  composition sensitivity over the whole energy range as the reconstruction bias and the statistical fluctuations are decreasing, which is shown in Figure \ref{Plt:RFReco} (right plot).

\section{Summary and Outlook}
The IceAct telescopes are compact Cherenkov telescope prototypes to measure cosmic ray air showers in the energy range from 10 TeV to a few PeV.
Two IceAct telescopes with 61 SiPM pixels are successfully operated at the South Pole since 2019.
The data acquisition systems were unified in 2020 to a TARGET-C based system representing the foreseen standard production telescope configuration. 
A drift corrected event-to-event synchronisation for the IceCube and IceAct events has been performed for a test data-set acquired in 2019. 
The resulting event selection method shows this method performs as expected for the single telescope events as well as stereo events.
A first analysis of the angular geometric event reconstruction with an image reconstruction using basic statistics (Hillas analysis) of IceAct events to the muon track analysis of the in-ice IceCube detector shows a good correlation between both components. The accuracy of the angular reconstruction of IceAct is estimated to be better than $2^\circ$.
A joined hybrid detector simulation of IceAct and IceCube/IceTop has been developed and the first results of the data/MC comparison are shown. These are  showing a decent agreement the results are expected to improve by increasing the simulated energy range and improving the simulation chain.
The combination of a simple statistical image analysis (Hillas parameters) with information of the other detector components shows promising composition sensitivity.
Currently multiple telescopes are under construction to be added to the current setup at the South Pole, increasing the total field-of-view to 36$^\circ$.
For future analyses we are working on an improved geometry reconstruction and a better energy resolution for pure telescope and hybrid events which will include the well monitored South Pole atmosphere. This will provide a better reconstruction of the primary particle leading to a better understanding of the mass composition of cosmic rays in the energy range from 10 TeV to a few PeV, which is at the South Pole only accessible with IceAct.

\bibliographystyle{ICRC}
\bibliography{references}



\clearpage
\section*{Full Author List: IceCube Collaboration}




\scriptsize
\noindent
R. Abbasi$^{17}$,
M. Ackermann$^{59}$,
J. Adams$^{18}$,
J. A. Aguilar$^{12}$,
M. Ahlers$^{22}$,
M. Ahrens$^{50}$,
C. Alispach$^{28}$,
A. A. Alves Jr.$^{31}$,
N. M. Amin$^{42}$,
R. An$^{14}$
% ...
\\

\noindent
% ...
$^{12}$ Universit{\'e} Libre de Bruxelles, Science Faculty CP230, B-1050 Brussels, Belgium \\
% ...
$^{14}$ Department of Physics and Laboratory for Particle Physics and Cosmology, Harvard University, Cambridge, MA 02138, USA \\
% ...
$^{17}$ Department of Physics, Loyola University Chicago, Chicago, IL 60660, USA \\
$^{18}$ Dept. of Physics and Astronomy, University of Canterbury, Private Bag 4800, Christchurch, New Zealand \\
% ...
$^{22}$ Niels Bohr Institute, University of Copenhagen, DK-2100 Copenhagen, Denmark \\
% ...
$^{28}$ D{\'e}partement de physique nucl{\'e}aire et corpusculaire, Universit{\'e} de Gen{\`e}ve, CH-1211 Gen{\`e}ve, Switzerland \\
% ...
$^{31}$ Karlsruhe Institute of Technology, Institute for Astroparticle Physics, D-76021 Karlsruhe, Germany  \\
% ...
$^{42}$ Bartol Research Institute and Dept. of Physics and Astronomy, University of Delaware, Newark, DE 19716, USA \\
% ...
$^{50}$ Oskar Klein Centre and Dept. of Physics, Stockholm University, SE-10691 Stockholm, Sweden \\
% ...
$^{59}$ DESY, D-15738 Zeuthen, Germany \\
% ...

\subsection*{Acknowledgements}

\noindent
USA {\textendash} U.S. National Science Foundation-Office of Polar Programs,
U.S. National Science Foundation-Physics Division,
U.S. National Science Foundation-EPSCoR,
Wisconsin Alumni Research Foundation,
Center for High Throughput Computing (CHTC) at the University of Wisconsin{\textendash}Madison,
Open Science Grid (OSG),
Extreme Science and Engineering Discovery Environment (XSEDE),
Frontera computing project at the Texas Advanced Computing Center,
U.S. Department of Energy-National Energy Research Scientific Computing Center,
Particle astrophysics research computing center at the University of Maryland,
Institute for Cyber-Enabled Research at Michigan State University,
and Astroparticle physics computational facility at Marquette University;
Belgium {\textendash} Funds for Scientific Research (FRS-FNRS and FWO),
FWO Odysseus and Big Science programmes,
and Belgian Federal Science Policy Office (Belspo);
Germany {\textendash} Bundesministerium f{\"u}r Bildung und Forschung (BMBF),
Deutsche Forschungsgemeinschaft (DFG),
Helmholtz Alliance for Astroparticle Physics (HAP),
Initiative and Networking Fund of the Helmholtz Association,
Deutsches Elektronen Synchrotron (DESY),
and High Performance Computing cluster of the RWTH Aachen;
Sweden {\textendash} Swedish Research Council,
Swedish Polar Research Secretariat,
Swedish National Infrastructure for Computing (SNIC),
and Knut and Alice Wallenberg Foundation;
Australia {\textendash} Australian Research Council;
Canada {\textendash} Natural Sciences and Engineering Research Council of Canada,
Calcul Qu{\'e}bec, Compute Ontario, Canada Foundation for Innovation, WestGrid, and Compute Canada;
Denmark {\textendash} Villum Fonden and Carlsberg Foundation;
New Zealand {\textendash} Marsden Fund;
Japan {\textendash} Japan Society for Promotion of Science (JSPS)
and Institute for Global Prominent Research (IGPR) of Chiba University;
Korea {\textendash} National Research Foundation of Korea (NRF);
Switzerland {\textendash} Swiss National Science Foundation (SNSF);
United Kingdom {\textendash} Department of Physics, University of Oxford.

\end{document}